\documentclass[manuscript]{aastex}
\usepackage{url}

\shorttitle{Evolutionary of ``black widow'' pulsars}

\shortauthors{O. G. Benvenuto, M. A. De Vito \& J. E. Horvath}

\begin{document}

\title{THE QUASI-ROCHE LOBE OVERFLOW STATE IN THE EVOLUTION OF CLOSE BINARY 
SYSTEMS CONTAINING A RADIO PULSAR}

\author{O. G. Benvenuto$^{1}$, M. A. De Vito$^{2}$}
\affil{Facultad de Ciencias Astron\'omicas y Geof\'\i sicas, Universidad 
Nacional de La Plata\\
and Instituto de Astrof\'\i sica de La Plata (IALP), CCT-CONICET-UNLP. Paseo 
del 
Bosque S/N (B1900FWA), La Plata, Argentina} \and
\author{J. E. Horvath$^{3}$}
\affil{Instituto de Astronomia, Geof\'\i sica e Ci\^encias Atmosf\'ericas, 
Universidade de S\~ao Paulo\\
R. do Mat\~ao 1226 (05508-090), Cidade Universit\'aria, S\~ao Paulo SP, Brazil}

\begin{abstract} We study the evolution of close binary systems formed by
a normal (solar  composition), intermediate mass donor star together with
a neutron star. We  consider models including irradiation feedback and
evaporation. These  non-standard ingredients deeply modify the mass
transfer stages of these  binaries. While models that neglect irradiation
feedback undergo continuous,  long standing mass transfer episodes,
models including these effect suffer a  number cycles of mass transfer
and detachment. During mass transfer the systems  should reveal
themselves as low-mass X-ray binaries (LMXBs), whereas when  detached
they behave as a binary radio pulsars. We show that at these stages 
irradiated models are in a Roche lobe overflow (RLOF) state or in a {\it 
quasi}-RLOF state. Quasi-RLOF stars have a radius slightly smaller than
its  Roche lobe. Remarkably, these conditions are attained for orbital
period and  donor mass values in the range corresponding to a family of
binary radio pulsars  known as ``redbacks''. Thus, redback companions
should be quasi-RLOF stars. We  show that the characteristics of the
redback system PSR~J1723-2837 are accounted  for by these models.

In each mass transfer cycle these systems should switch from LMXB to
binary  radio pulsar states with  a timescale of $\sim$million years.
However, there is  recent and fast growing evidence of systems switching
on far shorter, human  timescales. This should be related to
instabilities in the accretion disc  surrounding the neutron star 
and/or radio ejection, still to be included in the model having 
the  quasi-RLOF state as a general condition. \end{abstract}


\noindent$^{1}$ Member of the Carrera del Investigador  Cient\'{\i}fico of the 
Comisi\'on de Investigaciones
Cient\'{\i}ficas (CIC) de la Provincia de Buenos Aires, Argentina.\\ 
\email{obenvenu@fcaglp.unlp.edu.ar}

\noindent$^{2}$ Member of the Carrera del Investigador Cient\'{\i}fico,  
CONICET, Argentina. \\ \email{adevito@fcaglp.unlp.edu.ar}

\noindent$^{3}$ \email{foton@iag.usp.br}

\section{INTRODUCTION} \label{sec:introd}

%
%

Today it is usually accepted that millisecond pulsars (MSPs)  are old
neutron  stars (NSs) recycled in  close binary systems (CBSs)
\citep{1982Natur.300..728A}. In this  process, the companion (a low-mass
star) transfers mass and angular momentum to the NS during  a Roche Lobe
OverFlow (RLOF) episode, increasing the mass and rotation rate of the
NS.  During the mass transfer process the system is a copious source of
X-rays, generally (but  not only) detectable as a low-mass X-ray binary
(LMXB). For this type of systems,  the  standard models of CBSs predict a
long and stable episode of mass transfer, as a  consequence of the
evolution of the donor star, with a small number of RLOFs due to
thermonuclear  flashes (see, e.g., \citealt{1983ApJ...270..678W}; 
\citealt{1999A&A...350..928T}; \citealt{2002ApJ...565.1107P};
\citealt{2005MNRAS.362..891B}). As a final result, the  formation of a
MSP (with  a spin period $P_{rot} < 10$~ms) is expected. The most common
companion is a  low-mass helium white dwarf (HeWD), with characteristic 
mass values of $M_{2} \sim 0.2 - 0.3~M_{\odot}$\footnote{As usual, we
denote the quantities related to the NS and donor/companion star with
subscripts $NS$ and $2$, respectively.}. These objects are compact
remnants of CBSs with $P_{orb} > 1$~day (where $P_{orb}$ is the
orbital period of the system). Other physical phenomena that play a relevant 
role in the formation of MSPs are magnetic braking and gravitational radiation;
see, e.g., \citet{1993ARA&A..31...93V} for a review.

%
%

Standard models support the idea that MSPs descend from LMXBs, however
the  existence  of many isolated MSPs represent a serious difficulty to
the model. Nevertheless, \citet{1988Natur.333..237F}  discovered a MSP in
an eclipsing  binary system (PSR~1957+20). It has a spin period of 1.6~ms,
in orbit around a low-mass companion  of only $\sim 0.02~M_{\odot}$. Due
to the strong eclipses and the small mass of the  companion, it was
proposed that pulsar radiation is ablating the companion
\citep{1988Natur.333..832P}.  This is the representing object of a class
of  eclipsing binary systems, composed by a MSP and a very low-mass
companion ($M_{2} <  0.05~M_{\odot}$), in tight orbits ($P_{orb} < 1$~d), generally called ``black
widows''. The evaporation of the companion by the pulsar emission in CBSs
could be a process that allow recycled NSs to unbind their  companions
and finally appear as isolated MSPs.

%
%

It has been recently shown (\citealt{2014ApJ...786L...7B}; hereafter BDVH) that
evolutionary trajectories leading to the evaporating (black widows) {\it locus} 
cross a region in the $P_{orb}-M_{2}$ plane in which accretion-driven 
irradiation causes an 
evolution markedly different from the standard one leading to LMXBs. In fact, there is a 
wide zone 
populated by pulsar companions that standard models of close binary evolution 
cannot 
reproduce. It is precisely this stage that will be studied in this paper.

%
%

Sometime ago it has been proposed that the pressure due to radio pulsar
irradiation may preclude accretion onto the NS. This phenomenon is usually 
referred to as ``radio ejection'' (\citealt{1989ApJ...336..507R}; 
\citealt{2001ApJ...560L..71B}). If radio ejection were fully  effective in
inhibiting accretion, the results described in BDVH would be no  longer valid.
Also, evaporation may be relevant at stages earlier than those corresponding to
black widows \citep{2013ApJ...775...27C}.

%
%

Perhaps the most important non-standard effect in the early stages of CBSs is 
irradiation feedback
\citep{2004A&A...423..281B}. The accretion onto the NS releases X-rays, which 
are partially
absorbed by the companion and produce a partial blockage of the energy that 
emerges from
the donor star interior. As a non-trivial result, the mass transfer episode 
becomes {\it cyclic}
(see \citealt{2004A&A...423..281B}; \citealt{2012ApJ...753L..33B}).
Calculations (see BHDV) predict a  back-and-forth switch of the system between
LMXB and binary radio pulsar states. Irradiation swells the companion and the 
latter is filling, or
quasi filling its Roche lobe. The inclusion of this new ingredient
allowed us to perform more realistic calculations, and to reconcile theory with
observations.

%
%

Direct proof of the theoretical scenario for recycling old NSs in MSPs was found for first time in SAX~1808.4-3658. \citet{1998Natur.394..344W} reported coherent millisecond X-ray pulsations in the 
persistent flux of this CBS. The source will probably 
become a MSP when accretion turns off. The parameters ($P_{orb}= 2$~h, eccentricity $e < 5 
\times 10^{-4}$ and pulsar mass function $f_{NS} = 3.7789(2) \times 10^{-5}~M_{\odot}$;
\citealt{1998Natur.394..346C}) imply a companion mass of 
$M_{2}= 0.14~-~0.18~M_{\odot}$, if
the NS mass is in the range $M_{NS}= 1.35~-~2.0~M_{\odot}$. The orbital period 
($P_{orb} < 1$~d) and the mass  of the companion  ($0.1 < M_2 / M_{\odot} < 0.4$) place this 
system in the redback region \citep{2013IAUS..291..127R} in the plane $P_{orb}-M_{2}$.
\citet{2013MmSAI..84..117B} have proposed that this object is undergoing a
switch between accretion and MSP stages dominated by radio ejection.

%
%

A handful of systems placed in the redback region have shown clear signs of disc 
states and extended radius of the donor (although not always simultaneously). For example, 
the FIRST~J102347.67+003841.2 source with $P_{orb}= 4.75$~h \citep{2005AJ....130..759T} is 
coincident with the MSP PSR~J1023+0038
\citep{2009Sci...324.1411A}. The 2001 observations showed evidence of mass outflow, but
since 2002 a quiescent state followed. Later it has been found that for 
reasonable NS masses, the companion mass is
limited to $0.14\lesssim M_{2}/M_{\odot} \lesssim 0.42$ and likely fills its 
Roche lobe because of
the presence of ionized material. The recent (mid-2013) enhancement of UV and X-ray 
emission (\citealt{2013arXiv1311.7506S};  \citealt{2014ApJ...781L...3P}), together with 
the interruption of the radio pulsations, has been interpreted as evidence that
the LMXB phase returned \citep{2009Sci...324.1411A}. Since the system is
far from any globular cluster and the orbit is almost circular, it is reasonable to assume
that the pulsar has been recycled by the companion and alternates between two states
with the companion filling the Roche lobe. This system may be seen as
redback in the MSP phase and turn to LMXB phenomenology when accretion 
(re)starts. 

Redback seemed to be rare, but since 2009 many of them were discovered in the 
Galactic plane
\citep{2013IAUS..291..127R}, as well as in globular clusters  (see Pulsars in 
Globular Clusters, Freire's
homepage \footnote{http://www.naic.edu/~pfreire/GCpsr.html}). Currently, there 
are $\sim 7$ redbacks known in
the Galactic field, and $\sim 12$ in globular clusters.

%
%

Perhaps the best example of this class of objects, changing its state between 
accreting X-ray
MSP and radio pulsars state is precisely a system located in the globular 
cluster M28,
PSR~J1824-2452I (with $P_{rot}= 3.9$~ms and $P_{orb}=11$~h).
The system was detected switching from an accretion phase to a radio pulsar 
phase in the last decade
\citep{2013Natur.501..517P}. The change between rotation and accretion powered 
states in a few-days to
few-months mark the existence of a phenomenon that occurs on a timescale much 
shorter that the usual
long scale of the order of $\sim$~gigayears in these type of CBSs, or the 
pulsed mass transfer scale recently found in BDVH.

%
%

Further systems belonging to the redback family and displaying evidence for an 
extended
radius of the companion, even in quiescence, include PSR~J1723-2837. This is an 
eclipsing, $P_{rot}= 1.86$~ms
binary radio pulsar in an almost circular orbit with $P_{orb}=14.76$~h
discovered by  \citet{2004MNRAS.355..147F}. The eclipses suggested that the 
pulsar companion is a
non-degenerate, extended star. This has been confirmed by 
\citet{2013ApJ...776...20C}, who
identify the companion in the IR, optical, and UV bands.
The spectral analysis put the companion between G5 and K0 spectral types with 
an effective temperature
of 5000-6000~K and a surface gravity reasonably consistent with a main sequence 
star. These authors derived a mass ratio of $M_{NS}/M_{2}= 3.3 \pm 0.5$, giving a companion mass 
range of $M_{2}= 0.4-0.7~M_{\odot}$, 
and an orbital inclination angle between $30^{\circ}-41^{\circ}$,
assuming a pulsar mass in the range of $M_{NS}=1.4-2.0~M_{\odot}$. The derived radius 
of the companion is larger than expected, likely close to filling its
Roche lobe, i.e., a {\it quasi}-RLOF object.

An example of a quasi-RLOF object is PSR~J1740-5340. It is a binary MSP
($P_{rot}=$~3.65~ms) that has $P_{orb}=$~32.5~h, which falls outside the usually
considered range of values corresponding to redbacks. For this object 
\citet{2002ApJ...574..325B} proposed a possible binary progenitor that accounts
for the observational data of the companion and considers that it is on a radio
ejection stage. 

%
%

XSS~J12270-4859, on the other hand, experienced a transition from LMXB state to 
radio MSP (\citealt{2014arXiv1402.0765B}; \citealt{2014ATel.5890....1R}). This report 
confirm the idea 
that this pair belongs to the group formed by a low-mass quasi-RLOF object and 
a rotation-powered MSP, a
redback system in the current state of XSS~J12270-4859. 

Finally, we quote the 
cases of PSR~J1023+0038 
and XSS~J12270-4859, which experienced transitions in opposite directions. In 
the transition of
PSR~J1023+0038 from MSP to LMXB, it was detected an increase in the flux of 
$\gamma$-rays \citep{2013arXiv1311.7506S}. 
XSS~J12270-4859 showed a decrease of the $\gamma$-ray brightness (Tam et al 
2013), although less intense 
than the growth in the case of PSR~J1023+0038. The transition of PSR~J1023+0038 
occurs in
$\sim 2$~weeks, while the transition of XSS~J12270-4859 was in $\sim 5$~weeks. 
The process of back-and-forth
transitions seem to be fast.

%
%

Besides the eclipses and a non-degenerate low-mass companion, redbacks seems to 
have another
feature that identified them: the high filling factor of the Roche lobe by the 
donor star component
($\gtrsim 90\%$ in many cases, see \citealt{aspen} and referenced therein).

%
%

All the phenomenology described above can not be solely related to the 
evolution 
of the donor star,
because the latter has a typical timescale far greater. It seems more natural 
to 
interpret the changes between
LMXB and binary radio pulsar as due to instabilities of the accretion disc 
surrounding the NS. But for
these instabilities to occur the companion star must act as a donor. If we 
observe a binary MSP showing redback 
features, it would be hard to distinguish if it is visible because of disc 
instabilities or if it is detached
due to donor evolution. A population synthesis should be helpful to discern 
between these possible situations, 
because the time fraction of the cycle in which mass transfer occurs is small 
(see BDVH) and this is the only 
situation in which the pulsar should be detectable due to accretion disc 
instabilities (otherwise there is 
no accretion). If the scenario we are suggesting here is at least qualitatively 
correct, we should expect 
that redbacks should be more abundant than LMXBs with orbital periods and 
companion masses in the same range. 
Unfortunately, it seems that the number of already detected systems of these 
kinds prevent us to perform a 
solid statistical inference. In any case, the very existence of quasi-RLOF 
stars 
should be related to the irradiation feedback phenomenon.

%
%

It is the aim of this paper to discuss the quasi-RLOF state and also to show 
that the evolutionary models 
presented in BDVH provide a natural account of the characteristics of the 
redback system PSR~J1723-2837.

%
%

The remainder of this paper is organized as follows: In Section~\ref{sec:code} we
briefly describe the ingredients of our stellar code. In Section~\ref{sec:qrlof} 
we present
the quasi-RLOF state  in the evolution of CBSs containing
radio pulsars. In Section~\ref{sec:aplicando} we apply this novel evolutionary 
state to the
case of the available observations of the PSR~J1723-2837 redback system. In
Section~\ref{sec:disc} we discuss the meaning of the results presented in this 
paper and
finally, in Section~\ref{sec:concl} we give some concluding remarks.

\section{THE NUMERICAL CODE} \label{sec:code}

The results to be described below have been found employing our CBS evolutionary code
\citep{2003MNRAS.342...50B}  updated by the inclusion of irradiation 
feedback and evaporation. During a RLOF episode the code solves for the donor star 
structure, the instantaneous mass transfer rate $\dot{M}$, both masses and 
orbital semi axis {\it implicitly}. This method allows
for the computation of the mass transfer cycles quoted in Section~\ref{sec:introd} 
to be presented below in detail. In detached conditions, the code employs the standard Henyey technique.

We assumed that the NS  accretes a fraction $\beta$
of the transferred matter up to the Eddington critical rate
$\dot{M}_{Edd}\approx 2 \times 10^{-8} M_{\odot}/y$, while very
low  accretion rates ($\lesssim 1.3 \times 10^{-11}
M_{\odot}/y$) are  inhibited by the propeller mechanism
\citep{2008AIPC.1068...87R}. Angular momentum sinks due to
gravitational   radiation, magnetic braking and mass loss from the system
have been treated as  in \citet{2003MNRAS.342...50B}.

Irradiation feedback has been included following
\citet{1997A&AS..123..273H} who replace the usual relation between luminosity, 
radius and effective temperature $L= 4 \pi R_{2}^{2}\; \sigma\;
T_{eff}^{4}$ by

\begin{equation}
L= R_{2}^{2}\; \sigma\; T_{eff,0}^{4}\; \int_{0}^{2\pi} \int_{0}^{\pi} 
G(x(\theta,\phi))\; \sin{\theta} \; d\theta\; d\phi, \label{eq:irradiando}
\end{equation} 

\noindent where $\sigma$ is the Stefan-Boltzmann constant, 
$T_{eff,0}$ is the
effective  temperature of the non-illuminated part of the star, $G(x)=
\big(  T_{eff}(x)/T_{eff,0} \big)^{4} - x$ and $x= F_{irrad}/(\sigma\;
T_{eff,0}^{4})$,  where $F_{irrad}$ is the incident irradiating flux. 
$T_{eff}(x)$ is computed assuming that at
the deep enough layers, perturbations due to irradiation must vanish 
\citep{1985A&A...147..281V}. 

We assume that the NS acts as an isotropic {\it point source}, releasing an accretion
luminosity  $L_{acc}= G M_{NS} \dot{M}_{NS}/R_{NS}$ ($R_{NS}$, and
$\dot{M}_{NS}$  are the radius and accretion rate of the NS, 
respectively). The irradiation flux incident onto the
donor star, effectively participating in the
feedback phenomenon is $F_{irrad}= \alpha_{irrad} L_{acc}/ 4\pi  a^{2}$, where 
$\alpha_{irrad}$ is the fraction of the flux irradiated towards the donor that drives the phenomenon, and $a$ is the orbital radius. Because of the assumption of isotropy and the way it has been defined,
we have $0 \leq \alpha_{irrad} \leq 1$.

Because in the models to be presented below we assume that evaporation do not play a role we shall not detail the way we handled this phenomenon in our code. For further details we refer the reader to BDVH an 
references therein.

\section{THE QUASI-ROCHE LOBE OVERFLOW STATE} \label{sec:qrlof}

The occurrence of the quasi-RLOF state is easy to understand by performing a 
close examination of the evolution of the donor mass when irradiation feedback is considered. Before the 
onset of a RLOF, we assume there is no accretion onto the NS; so, there is no irradiation 
incident onto the donor star. Just after the beginning of the RLOF, irradiation feedback 
forces the donor star to have an instantaneous mass transfer rate $\dot{M}$ that overflows the value it would 
have had if irradiation were neglected. Let us define $\Delta\dot{M}\equiv max(\dot{M} -
\dot{M}_{nif})$ the maximum mass loss rate of irradiated models minus the
$\dot{M}_{nif}$ value it would have had without irradiation feedback. The 
general trend is that the stronger the irradiation regime, the larger $\Delta\dot{M}$.

In order to understand the quasi-RLOF state, for the moment let us assume that 
the donor star fulfills the conditions in which undergoes cyclic mass transfer for all values of
$\alpha_{irrad}>0$.

Initially, irradiation feedback leads to $\Delta\dot{M}>0$. After a time of the 
order of the Kelvin-Helmholtz timescale $\tau_{KH}$ (where $\tau_{KH}= G\; 
M^{2}_{2}/(R_{2}\;  L_{2})$, $G$
is the gravitational constant, and $R_{2}$, and $L_{2}$ are the radius and
luminosity of the  donor star, respectively) the star relaxes, damping the 
perturbation due to irradiation. Then, the donor star contracts because it has a mass {\it lower} 
than it would need to stay in a semi-detached state. Thus, the RLOF ends. While the 
stellar envelope reacts this way, the deep interior has a chemical profile evolving due
to nuclear reactions and mixing. Very often, this makes the star to swell 
enough, leading to the occurrence of another RLOF on a nuclear timescale. This is the very 
reason why stars may undergo a cyclic behavior.  The time the star needs to spend to reach 
another RLOF conditions (if any) depends on  the value of $\Delta\dot{M}$. If 
irradiation feedback is weak (e.g., the case of $\alpha_{irrad}$= 0.01) $\Delta\dot{M}$ is 
small and the star needs to evolve its internal chemical profile very little to RLOF 
again. Therefore, the average duration of the mass transfer cycles is rather short, 
leading to the occurrence of a large number of cycles. On the contrary, if irradiation 
feedback is stronger (e.g., up to $\alpha_{irrad}$= 1.00), $\Delta\dot{M}$  will be larger, 
the star will need more time to evolve to swell enough to RLOF again, mass transfer cycles 
will be longer and the star will undergo a much lower number of cycles.

\begin{figure} \begin{center}
\includegraphics[scale=.50,angle=0]{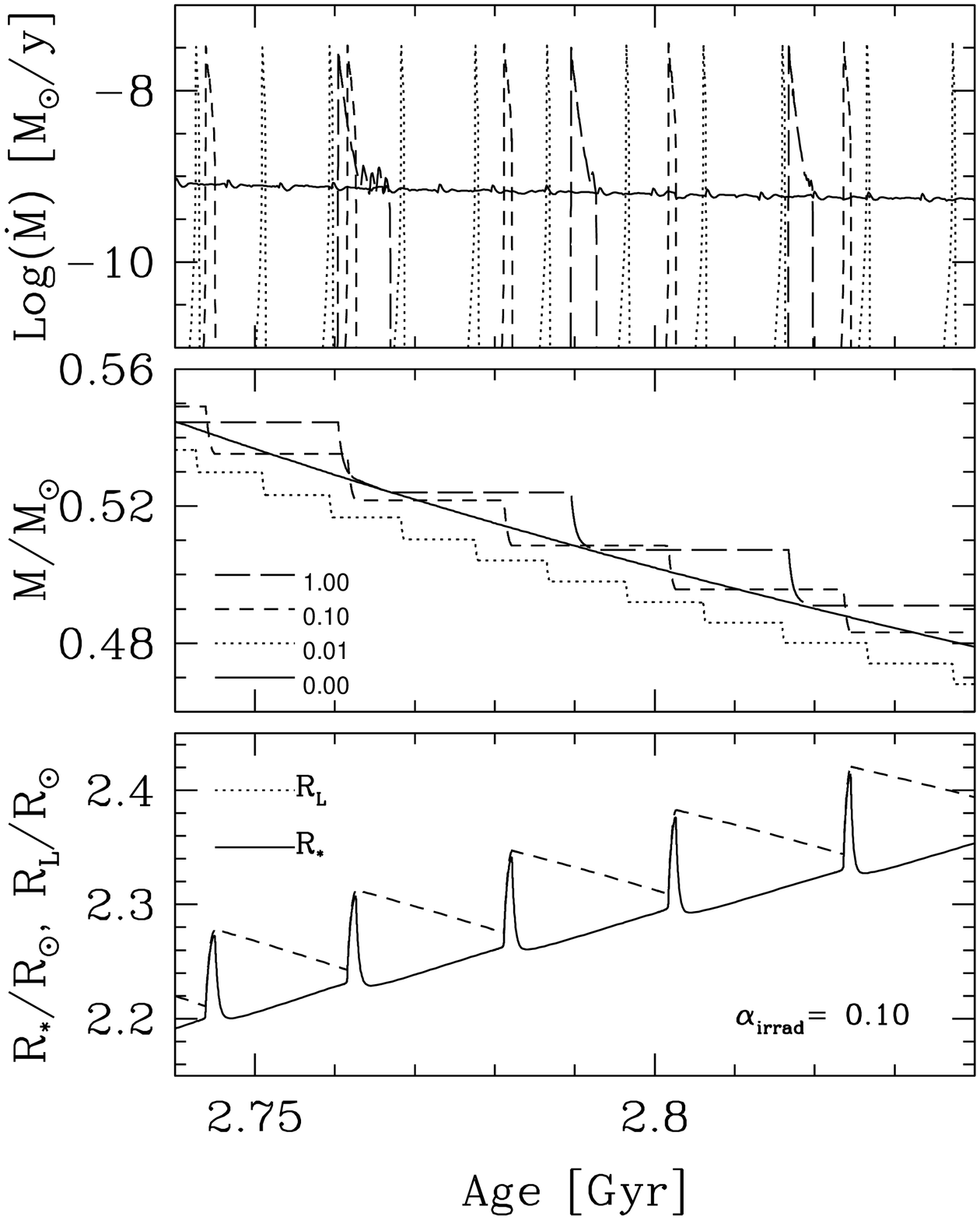}
\caption{Evolution of a donor star that initially has a mass 1.50~$M_{\odot}$, 
together with a
NS of 1.4~$M_{\odot}$, ($\beta$=0.5) on a 1~day orbital period, neglecting 
irradiation and also considering it with $\alpha_{irrad}= 0.01, 0.10$, and 
$1.00$.  Upper panel shows the mass transfer cycles while middle panel shows 
the 
corresponding evolution of the donor mass. Lower panel shows the evolution of 
the radius of the donor star and the equivalent Roche lobe. There, for the sake 
of clarity, we only considered the case of $\alpha_{irrad}= 0.10$. It is 
noticeable that after detachments, the star is in the quasi-RLOF state.
\label{fig:qrlof}} \end{center}
\end{figure}

The qualitative description given above can be analyzed with the help of
Fig.~\ref{fig:qrlof}. There we compare the evolution of the donor star for 
systems that initially had a normal star of 1.50~$M_{\odot}$, a ``canonical'' NS of 
1.4~$M_{\odot}$, and 1~day orbital period. We kept fixed $\beta=0.5$ and, in order to explore the 
whole parameter space of irradiation, we considered values of $\alpha_{irrad}= 
0.00, 0.01, 0.10$, and $1.00$. Here we have modeled the 
long timescale evolution of the systems.

In the upper and middle panels of Fig.~\ref{fig:qrlof} we display the mass transfer 
rate and the mass for these systems, complying with the behavior described above. While 
the model without irradiation shows a smooth evolution, irradiated models follow a 
step-like behavior in the mass. These steps are smaller the weaker the irradiation strength.

In the lower panel of Fig.~\ref{fig:qrlof} we show the evolution of the radius 
of the equivalent Roche lobe
against the stellar radius for the case of $\alpha_{irrad}= 0.10$. It is 
apparent that these radii are
almost the same during RLOFs (the difference of these radii is very small, 
being of the order of the pressure scale height
$H_{p}=dr/d\log{P}$ at the photosphere), while after detachment the star 
becomes slightly smaller than 
its associated Roche lobe. The star detaches because it has lost more mass than 
the amount it would if 
irradiation feedback were absent. In the latter conditions, the star would be 
on  a RLOF continuously. 
This represents a rather small perturbation to the stellar structure. Thus, 
there is no possibility for 
the donor star other than to have a radius slightly smaller than its Roche 
lobe. The donor star must settle on a {\it quasi-RLOF state}.

These arguments are quite general in the sense that if an irradiated donor star 
undergoes cyclic mass transfer,
it must be in the RLOF or in the quasi-RLOF state. This provides a natural 
explanation to the occurrence of
redback pulsars. Redbacks are a rapidly growing family of eclipsing binary 
systems containing a MSP together 
with a donor star with a mass $\sim 0.10~M_{\odot}$ and orbital periods from 0.1 
to 1.0~days. As described in BDVH, some redbacks evolve to the black widow state while 
others  give rise to the formation
of low-mass, HeWD-pulsar system. Our models support the idea that all redbacks 
{\it must} have a
non-degenerate companion in the quasi-RLOF state. This may also be the case for 
more massive objects usually
interpreted as carbon-oxygen~WD companions to pulsars (see Fig.~3 of  BDVH).

The very existence of the quasi-RLOF state indicates that pulsars may be 
observed accompanied not only by WDs
but also by non-degenerate companions (in fact, the deep interior of the donor 
companion should be at
densities high enough for the electrons to become semi-degenerate, but this is 
not the case of the extended
envelope). It also proves the relevance of considering irradiation feedback in 
CBS evolution.

\section{APPLICATION TO THE CASE OF THE PSR~J1723-2837 REDBACK SYSTEM} 
\label{sec:aplicando}

\begin{figure} \begin{center}
\includegraphics[scale=.50,angle=270]{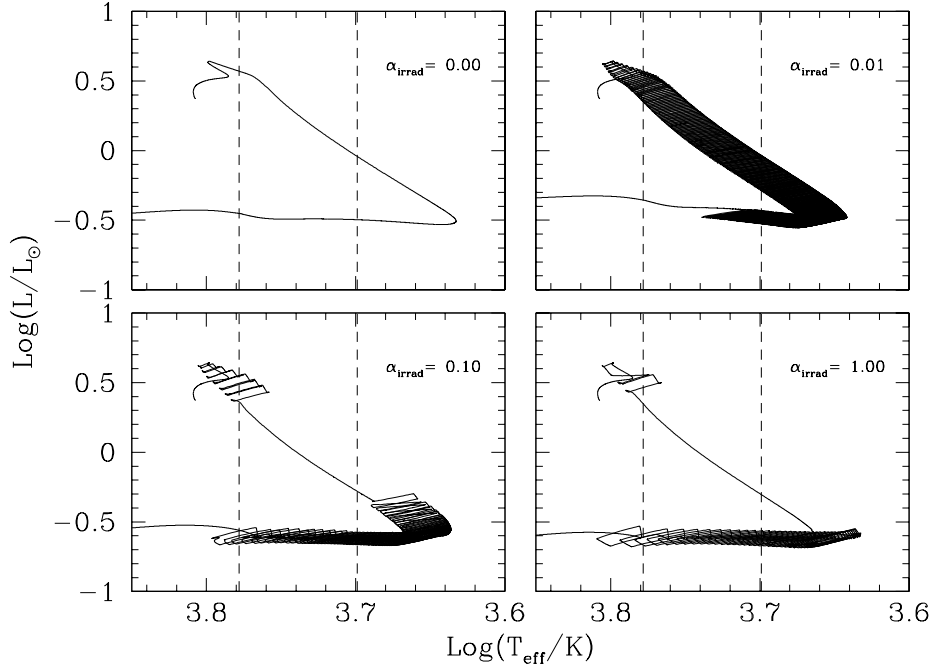} \caption{Evolutionary tracks 
for the donor star of a system
formed by a normal, solar composition donor star of 1.25~$M_{\odot}$ evolving 
on 
a CBS together with
a 1.4~$M_{\odot}$ mass NS on an 0.75~day orbit. Each panel correspond to 
different irradiation feedback
regimes indicated with the corresponding value of $\alpha_{irrad}$. The 
horizontal axis
corresponds to the effective temperature of the {\it non-}irradiated portion of 
the stellar surface.
Vertical dashed lines indicate the range of effective temperatures for the 
donor 
star in PSR~J1723-2837
system \citep{2013ApJ...776...20C}. \label{fig:hrd}} \end{center}  \end{figure}

\begin{figure} \begin{center}
\includegraphics[scale=.50,angle=270]{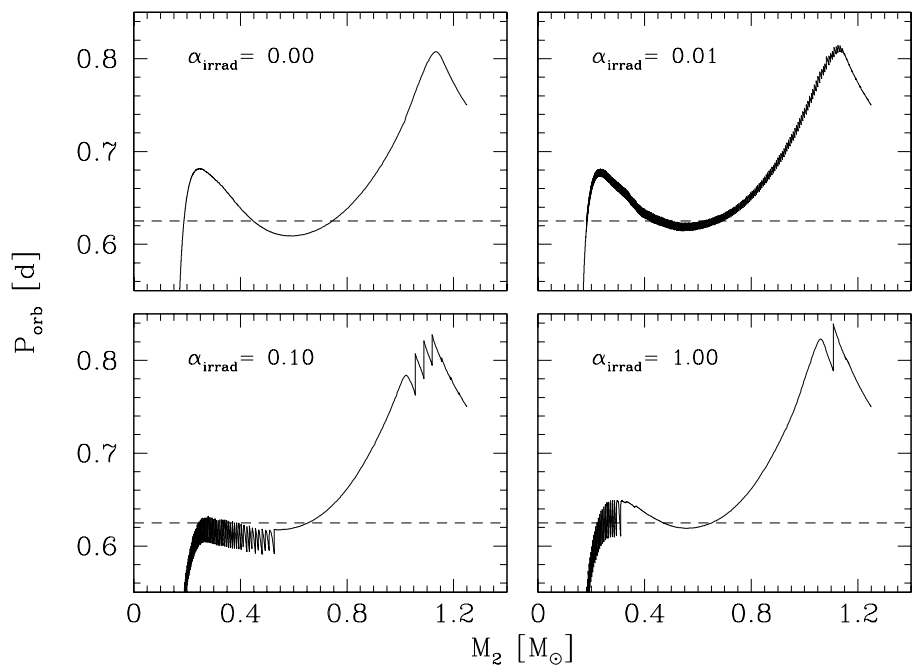} \caption{Donor mass vs. 
orbital period for the
systems included in Figure~\ref{fig:hrd}. The orbital period corresponding to 
PSR~J1723-2837
system is denoted with a horizontal dashed line. Notice that irradiated models 
reach the observed
period value for several times while the non-irradiated one provides three 
values for the donor mass.
\label{fig:perimasa}} \end{center}  \end{figure}

\begin{figure} \begin{center}
\includegraphics[scale=.50,angle=270]{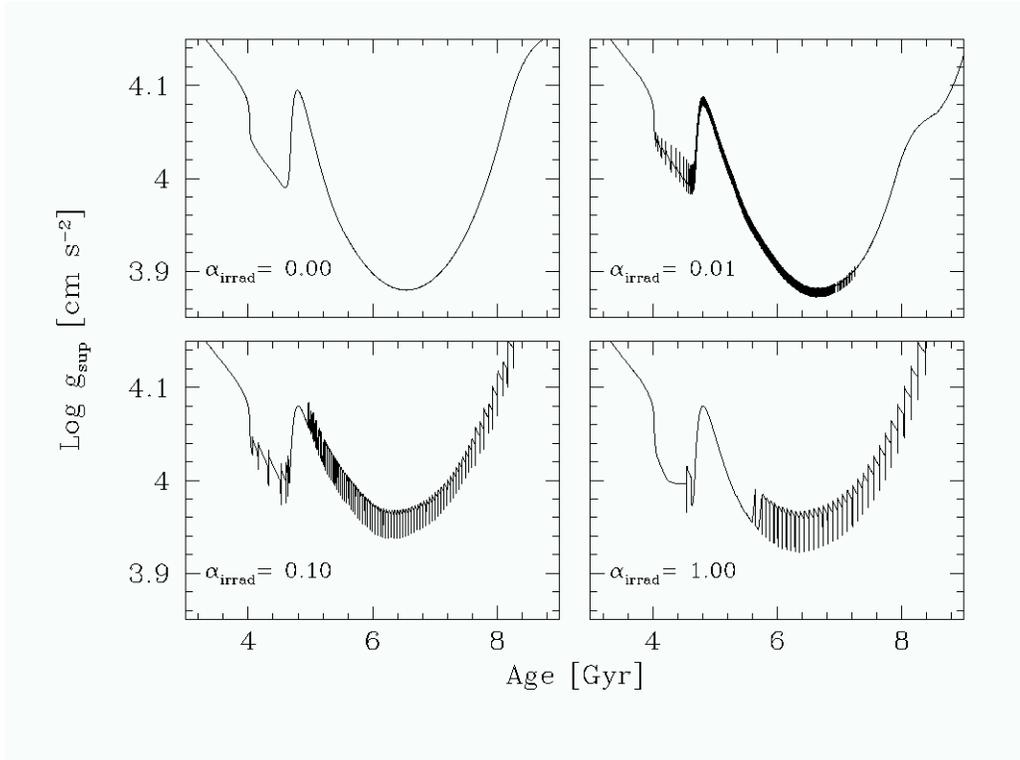} \caption{Surface gravity as 
a 
function of time for the
systems included in Figure~\ref{fig:hrd}. These surface gravity values are 
similar to those expected for a
main sequence star, as suggested by \citet{2013ApJ...776...20C} for the donor 
star of PSR~J1723-2837 system.
\label{fig:grave}} \end{center}  \end{figure}

\begin{figure} \begin{center}
\includegraphics[scale=.50,angle=270]{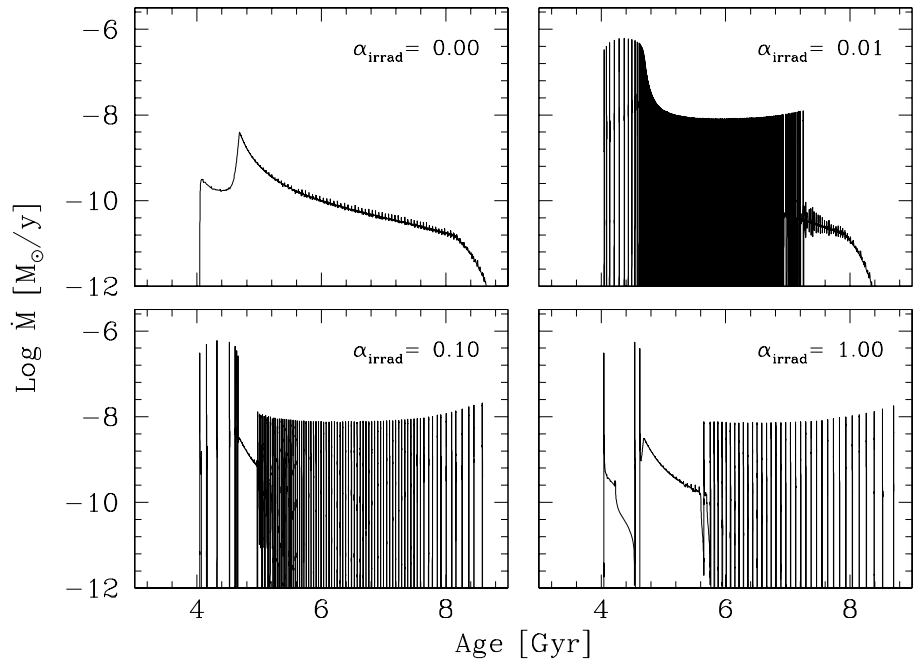} \caption{Mass transfer rate 
for the systems included in Figure~\ref{fig:hrd}.
\label{fig:mdot}} \end{center}  \end{figure}

\begin{figure} \begin{center}
\includegraphics[scale=.50,angle=0]{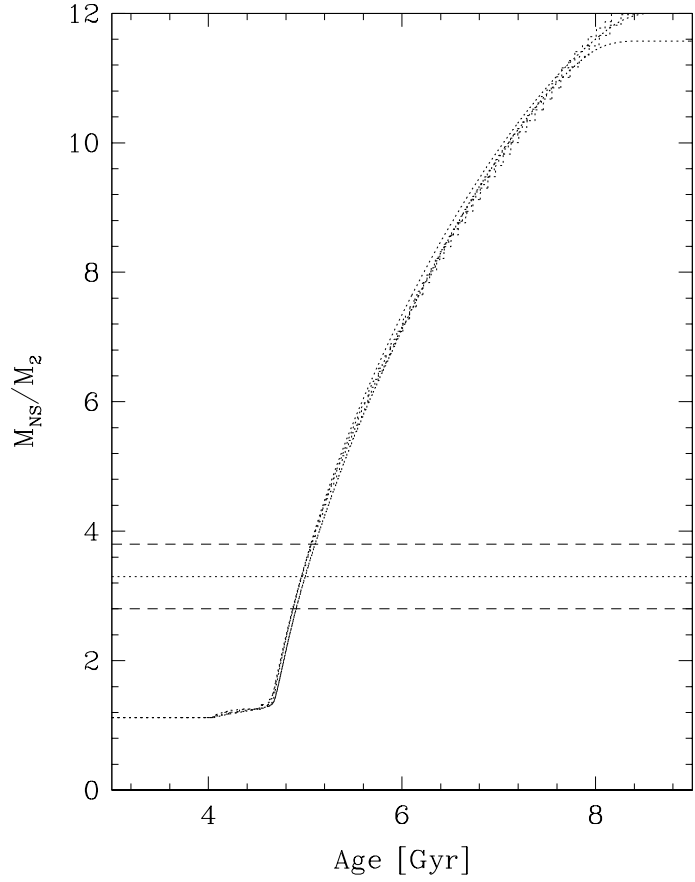} \caption{The ratio of the NS to 
donor mass as a function of time for the systems included in 
Figure~\ref{fig:hrd}. The mass ratio observed in  PSR~J1723-2837 system is 
denoted with a dotted line whereas its uncertainty is depicted with horizontal 
dashed lines. The models considered provide the correct mass ratio at an age of 
$\sim 4.9$~Gyr. \label{fig:qmasas}} \end{center}  \end{figure}

\begin{figure} \begin{center}
\includegraphics[scale=.50,angle=270]{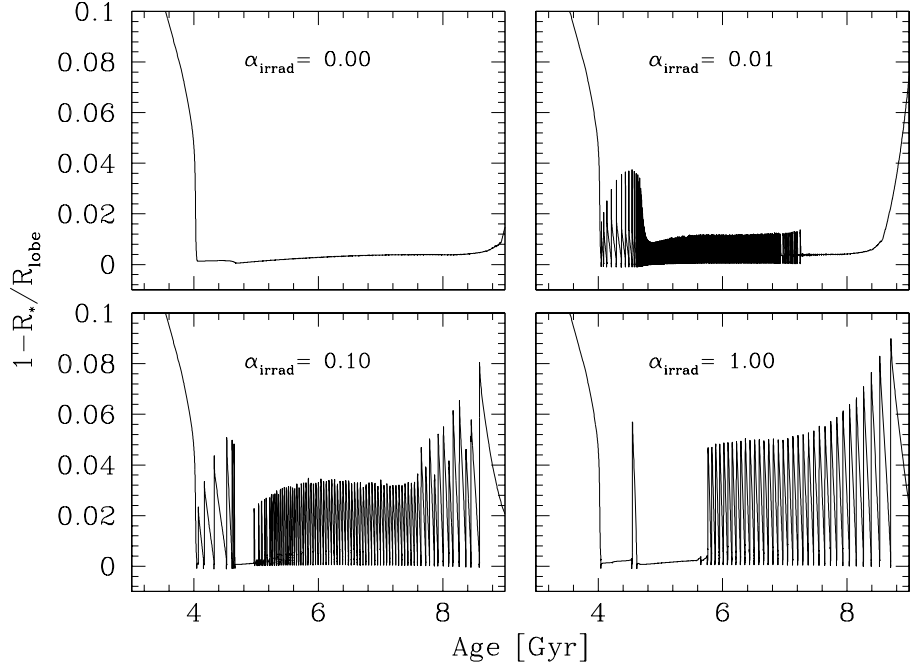} \caption{The unfilled 
fraction of the Roche lobe as a function of time for the systems included in 
Figure~\ref{fig:hrd}. Non-irradiated model has a filled Roche lobe for most of 
the time period depicted here (for ages $\geq 4$~Gyr) whereas irradiated models 
at detached conditions have a radius that differ from that of the Roche lobe at 
most in few percents. The donor star is in the quasi-RLOF 
state.\label{fig:llenado}} \end{center}  \end{figure}

As stated above, the system containing PSR~J1723-2837 is one of the best 
observed cases to be explained as a prototype.
In order to interpret the available observations of PSR~J1723-2837 in the 
framework of an evolutionary model we have searched among the set of calculations 
presented in BDVH. We found that the CBSs formed by a 
normal, solar composition 1.25~$M_{\odot}$ star together with a NS of 1.40~$M_{\odot}$ on a 0.75~day 
circular orbit are suitable for such purpose. We shall consider models with and
without irradiation feedback. With the 
purpose of exploring the whole parameter space of 
irradiation we considered four values of the coupling constant
$\alpha_{irrad}=$~0.00, 0.01, 0.10, and
1.00. The comparison of the available data on PSR~J1723-2837 and the 
evolutionary tracks will be performed with
the help of Figures~\ref{fig:hrd}-\ref{fig:llenado}

In Figure~\ref{fig:hrd} we present the evolutionary tracks in the HR diagram 
for each value of $\alpha_{irrad}$, 
where we indicate the range of effective temperatures compatible with the 
spectroscopic observations corresponding 
to the donor star presented by \citet{2013ApJ...776...20C}. There are two 
epochs 
compatible with observations, 
one before bending bluewards and another after it. In both cases, the star has 
surface conditions similar to 
those of main sequence stars, but its internal structure is quite different.

Figure~\ref{fig:perimasa} depicts the donor mass vs. orbital period 
relationship 
for the considered systems. 
The non-irradiated model reaches the observed orbital period of PSR~J1723-2837 
in three epochs while irradiated
models fulfill this condition several times. In all cases, the mass range at 
which the correct period is reached 
corresponds to the donor mass range given by \citet{2013ApJ...776...20C}. The 
evolution of the surface gravity 
is shown in Figure~\ref{fig:grave}. Again, these values are similar to those of 
main sequence stars, as it can be expected
by the results included in Figure~\ref{fig:hrd}. The mass transfer rate for 
these systems is given in
Figure~\ref{fig:mdot}. Models considering irradiation feedback are consistent 
with 
the observed phenomenology, since we do not observe a LMXB at all times.

Figure~\ref{fig:qmasas} denotes the mass ratio of the components of the systems 
considered above. The observed 
value of $M_{NS}/M_{2}= 3.3 \pm 0.5$ is attained at an age of $\sim 4.9$~Gyr. 
Finally, in Figure~\ref{fig:llenado} 
we present the unfilled fraction of
the Roche lobe as a function of time. Irradiated, detached models have radii 
smaller than the corresponding Roche 
lobe in at most a few percent. Clearly, the donor star is in the quasi-RLOF 
state as indicated by observations.

In the frame of our models, a CBS capable to account for the properties of 
the PSR~J1723-2837 system has to undergo cyclic mass transfer episodes. If, on 
the contrary,
mass transfer were continuous, we would observe the system as a LMXB. Our models 
provide the observed mass ratio of PSR~J1723-2837 system for an age of $\sim 
4.9$~Gyr, which correspond to the portion of the evolutionary track presented in 
Figure~\ref{fig:hrd} before bending bluewards. At 
this age, the models with $\alpha_{irrad}=$~0.00, 0.10, and 1.00 do not undergo 
cyclic mass transfer, meanwhile that with
$\alpha_{irrad}=$~0.01 does. Thus, low-irradiation regimes provide a 
satisfactory account of the properties
observed of the donor star in PSR~J1723-2837 system.

\section{DISCUSSION} \label{sec:disc}

In the previous section we have discussed the relevance of evolutionary tracks 
including irradiation feedback to the case of the redback system PSR~J1723-2837. 
We found that the observations reported by \citet{2013ApJ...776...20C} are 
consistent with  the theoretical picture with irradiation. Moreover, the 
analysis presented in Section~\ref{sec:aplicando} indicates that the quasi-RLOF 
state is reached by actual binary systems. In view of the generality of the 
arguments presented above, we suggest that essentially all redbacks should have 
quasi-RLOF companions. In other words, that redbacks are the observational 
manifestation of the quasi-RLOF state. This is supported by the results 
presented in \citet{aspen}, particularly the data related to the Roche lobe filling 
factors.

In this study we have considered a single value for the parameter 
$\beta$~(=0.5). This quantity is the fraction
of the matter the NS is assumed to accrete ($\dot{M}_{NS}= min[-\beta 
\dot{M}_{2}, \dot{M}_{Edd}]$). While it is 
well-known that the evolution of donor stars is largely independent of $\beta$, 
the mass of the NS is sensitively affected by this choice. There have been some 
studies indicating that the value of $\beta$ may be lower in some cases (see, 
e.g., \citealt{2006MNRAS.366.1520B} for the case of PSR~J1713+0747). If the NS 
mass growth were slower, this would modify the results presented in 
Figure~\ref{fig:qmasas}; and a good fit to PSR~J1723-2837 system would be 
provided by different initial conditions. In any case we should remark that the 
initial NS mass value is, to a large extent, also uncertain. Evidently there is 
some degeneracy for the initial parameters of these binary systems and thus, we 
are {\it not} claiming that the evolutionary tracks we considered here in 
Section~\ref{sec:aplicando} are the only possible ones. However, in finding the 
results presented in this paper we have not performed any fine tuning of the 
parameters of the evolutionary model, and yet found very suitable calculations 
on a previously computed set. Thus, for different values of $\beta$ other 
consistent solutions should exist as well.

Some redback systems  (PSR~J1023+0038 and XSS~J12270-4859) show evidence of
surrounding material \citep{2014arXiv1406.2384L} while others do not. In the
framework of our models, they should correspond to the RLOF and quasi-RLOF
states, respectively. Consequently, redbacks with surrounding material reveal as
pulsars not due to the long term evolution resulting from our calculations but
due to some physical processes (radio ejection and/or accretion disk
instabilities) still to be included in the models. The very existence of
redbacks with a clean environment shows that irradiation feedback plays a key
role in accounting for the evolution of these objects. If there were no clean
redback we could interpret that standard binary evolution coupled to radio
ejection and/or accretion disk instabilities may be enough to account for
observations. However, this seems not to be the case. 

The evolutionary status of redbacks has been recently addressed by 
\citet{2013ApJ...775...27C} employing models including evaporation. They state that
the evolution leading to redbacks or black widows is determined by the efﬁciency
of the irradiation process leading to mass loss via evaporation. They claim that
redback systems do not evolve into black widow ones. This scenario is deeply
different from the one described in BDVH, especially because
\citet{2013ApJ...775...27C} do not find some stage similar to what we called
quasi-RLOF state while in the frame of our calculations we interpret that all
redbacks correspond to this state. Moreover, we found that some redbacks evolve to
black widows (see BDVH for further details).

\section{CONCLUSIONS} \label{sec:concl}

We have discussed the occurrence of the quasi-RLOF state originally discussed 
in BDVH and applied it to the 
well-studied redback system PSR~J1723-2837 \citep{2013ApJ...776...20C}. We 
found that low-irradiation models 
describe very well the properties of this binary system, suggesting that the 
quasi-RLOF actually takes place along binary evolution. This also supports the 
relevance of irradiation feedback for this kind of binary systems. It is likely 
that the phenomenology observed in PSR~J1723-2837 system and in those quoted in 
the Introduction of this paper (Section~\ref{sec:introd}) is representative of 
the whole class of redbacks.

Taken at face value, on human timescales, our models are able to account for 
the switch of a given system from 
LMXB to binary radio pulsar, or in the opposite direction, only once. Detecting 
such a transition in the frame 
of our evolutionary models has an extremely low probability and would be very 
unlikely. However, in some systems (see Section~\ref{sec:introd}) this switch 
has been observed from one state to the other and then back to the original. 
Therefore, a short ($\sim$~1~yr) timescale is clearly present in these systems and 
does not stem from our theoretical framework. It seems natural to consider that 
the main ingredients to be included in the model to find such a short timescale 
for the switching behavior should be associated to the accretion disc 
surrounding the NS. It is known that accretion discs may undergo instabilities 
on short timescales (the disc instability model; see, e.g., 
\citealt{2001NewAR..45..449L}; \citealt{2012MNRAS.424.1991C}) that may fall in 
the required range. 

Recently, it has been possible to measure the orbital period change of
SAX~1808.4-3658 (\citealt{2008MNRAS.389.1851D}; \citealt{2009A&A...496L..17B}). The
reported value of $\dot{P}_{orb}= (3.85 \pm 0.15)\times 10^{-12}\;s s^{-1}$ is far
larger the value expected for conservative ($\beta=1$) binary evolution. These
authors interpret it as evidence for a strong mass loss from the system driven by
radio ejection. While this may correspond to the case of SAX~1808.4-3658, it is
worth to remark that the radiation pressure of a radio pulsar does not necessarily
destroys a surrounding accretion disk \citep{2005ApJ...620..390E}.

Finding whether radio ejection and/or disc instability model are able to  account
for the observations described in Section~\ref{sec:introd} remains an  open problem
and deserves further investigations. Despite this, we consider that the framework
for the occurrence of these complex phenomena is given by the long term binary
stellar evolution described in BHDV as well as in the present paper.


\clearpage

%
%

\clearpage
\end{document}